\documentclass[aps,noshowpacs,prl,letterpaper,twocolumn,nofootinbib,floatfix]{revtex4}
\usepackage{graphicx,color}
\usepackage{dcolumn}
\usepackage{bm}

\begin{document}
\title{Single-Base DNA Discrimination via Transverse Ionic Transport}

\author{James Wilson}
\author{M. Di Ventra}

\begin{abstract}
We suggest to discriminate single DNA bases via transverse {\it ionic} transport, namely by detecting the ionic current that flows in a channel while a single-stranded DNA is driven through an intersecting nanochannel. Our all-atom molecular dynamics simulations indeed show that the ionic currents of the four bases are {\it statistically} distinct, thus offering another possible approach to sequence DNA.
\end{abstract}
\maketitle
\section{Introduction}
Measuring transverse {\it electronic} currents within a nanochannel to differentiate between DNA/RNA bases \cite{lagerqvist:2006,zwolak:2005,krems:2009} is a promising new approach for sequencing DNA fast and at low cost. It consists of feeding a single-stranded DNA molecule through a channel equipped with nanometer-scale electrodes able to differentiate (via transverse electrical current) the electronic structure of the various bases as they pass by \cite{lagerqvist:2006,zwolak:2005,krems:2009}. This approach has been recently realized in various experiments \cite{ohshiro:2012,ivanov:2010,liang:2008,chang:2010,tsutsui:2011}.

On the other hand, ionic current through a nanopore has been known for over a decade to be useful in detecting when a DNA translocation event has occurred \cite{kasianowicz:1996, deamer:2002,merchant:2010,garaj:2010,schneider:2010,howorka:2009}, and has even been able to give some information about the sequence \cite{derrington:2010,manrao:2012}. However, within a nanochannel, the ionic current blockaded by an arbitrary strand of DNA is a non-trivial convolution of a large number of blockade events from different bases \cite{branton:2008}, and as such it is difficult to sequence at the single base level with this physical mechanism, unless the bases are fed one at a time through the opening.

Recent advances in the fabrication of ionic nanochannels \cite{menard:2012} have shown that it is possible to use a pair of intersecting nanochannels to detect the transport of DNA. In that particular experiment a double stranded DNA has been translocated through one channel, while the ionic current flowing through the second, transverse, channel is modulated based on the presence or absence of DNA at the intersection. So far, fabrication techniques have only realized nanochannel widths of about $20$ to $30$ nanometers, still too large to be able to achieve single-base resolution, if at all possible. In fact, the linear dimension of an individual base is on the order of $1$nm, so in order to distinguish a single base, channels of widths comparable to or less than $2$nm are required. Even if those were possible to make, it is not at all obvious whether single-base discrimination would be achievable with transverse ionic transport.

\begin{figure}[t]
\centering
\includegraphics[width=0.942\columnwidth]{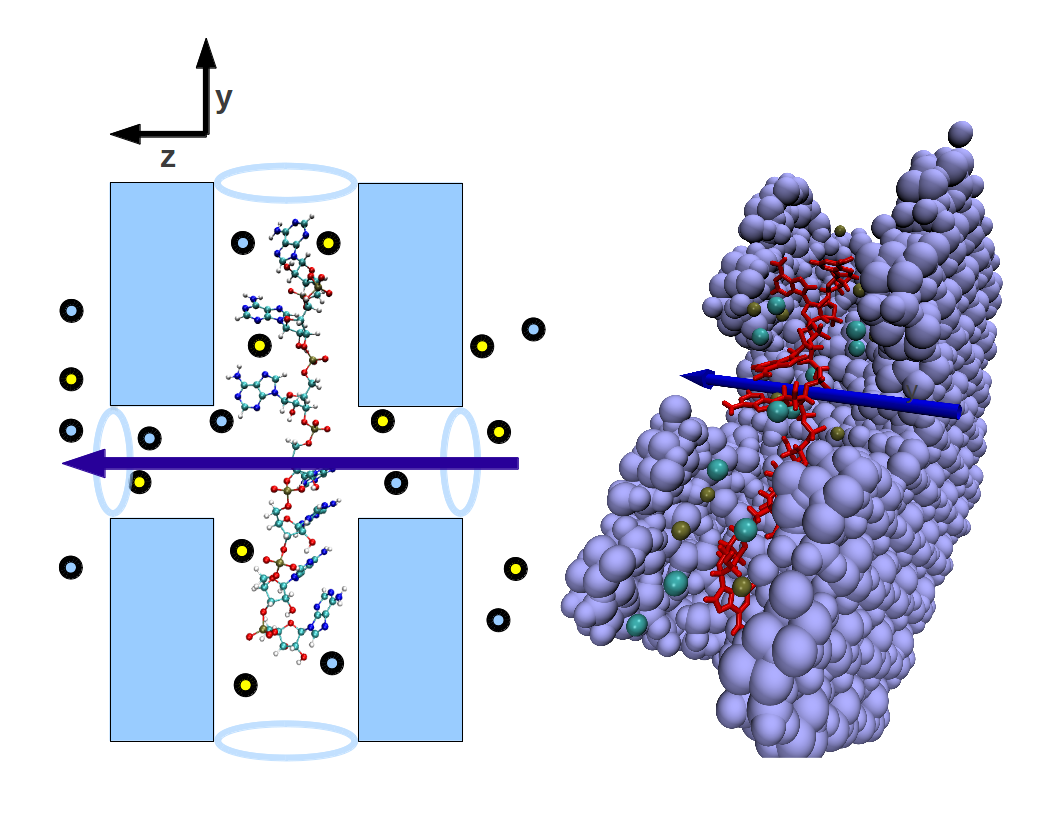}
\caption{\label{fig:Diagram}(Color online) Schematic of two intersecting nanochannels. A strand of DNA lies along the longitudinal (vertical) pore. The ionic current flows in the transverse (horizontal) direction symbolized by the arrow.
}
\end{figure}

Inspired by these experimental advances, and the possibility to realize devices with intersecting channels as those mentioned above, in this work we seek to determine whether changes in {\it transverse ionic conductance} are sufficient to discriminate the different DNA bases, were intersecting nanochannels of width comparable to or less than $2$nm fabricated. To be more specific, the set up we have in mind is illustrated in Fig.~\ref{fig:Diagram}. We place a strand of poly(dX)$_{7}$ where X is one of the bases $A$, $C$, $G$, $T$ in the pore. To optimize the simulation time required, the single-stranded DNA (ss-DNA) is already placed inside of a Si$_3$N$_4$ nanopore of diameter of $1.8$nm that runs along the $y$-axis. The simulation box has a regular hexagonal shape with an in-radius of $3.3$nm in the $x$-$y$  plane to correspond with hexagonal periodic boundary conditions. Intersecting this, a transverse pore of diameter $1.4$nm extends in the $z$-axis direction. The membrane is $2.8$nm thick in the $z$ direction and the simulation box, which contains water stacked in the $z$ direction, has rectangular periodicity in the $z$ direction.
In addition, an atom in the backbone of each of the outer two bases in the ss-DNA is fixed in space during the simulation. This allows the bases that are near the intersection of the pores to move without any constraint, but does not allow the DNA to move out of the pore or to fold up. Moreover, after each simulation run we vary the position of the two outermost bases along the $y$ axis by 0.1\AA, while also keeping the distance between the above two points fixed. This way the ionic currents we detect correspond to different configurations of the single bases, facing the transverse channel. In fact, bases were found to shift several angstroms during the simulation while the end points remained fixed.
\section{Simulation}

An electric field is then imposed in the $z$-direction causing ions to flow through the transverse channel in the Si$_3$N$_4$. As has been discussed in \cite{lagerqvist:2006} the electric field in the longitudinal and transverse directions can be independently manipulated to allow the DNA to translocate through the pore slowly, yet still have sufficient time to make measurements. Many measurements of current for each base can then be made, allowing a current \emph{distribution} to be built for each base as it passes the channel intersection.

The system is simulated using the molecular dynamics (MD) package NAMD \cite{phillips:2005}. It is solvated with $5$nm of water on each side of the channel in the $z$-axis and K and Cl ions are added such that the molarity of the system is $2$M~\footnote{The choice of a larger molarity than the typical 1M is again to reduce the already demanding computational requirements.}. We use periodic boundary conditions all around.  The static energy is minimized and then the system is brought up to room temperature. Next, a Langevin piston is used to equilibrate at room temperature and $1$ atmosphere of pressure in the NPT ensemble. Finally the system is run in the NVT ensemble with a Langevin damping term to keep the temperature steady. An electric field is applied transverse to the DNA such that the voltage drop across the cell in the $z$ direction is $0.5$V.  The first $2$ns are used as a further equilibration with the same conditions as the production run. During this first $2$ns the system approaches a steady-state current (see also below).

In the previous work covering transverse \emph{electronic} transport \cite{lagerqvist:2006,krems:2009}, the current was calculated using a single-particle scattering approach \cite{diventra:2008}. The molecular dynamics simulated the structural fluctuations inherent in the water-pore system, and these fluctuations caused the variations in the current. That study was possible because the timescales in which the electronic transport occurred (due to tunneling) are much shorter than the timescales of the structural fluctuations. Unfortunately, the ionic transport timescales are much longer, so we must simulate many nanoseconds to correctly capture the fluctuations of the ionic current. For each base, 31 simulations were run. In all, we have simulated $1736$ns of MD, with each simulation run of 14ns. The computational resources required to characterize the current distributions were significantly greater than in the electronic transport study. To accomplish this, we made use of a cluster of 30 dual-core nodes and 10 eight-core nodes. We also used computing time from the Open Science Grid project \cite{osg:2007}, scavenging time whenever resources were available.
\section{Results}
\begin{figure}[t]
\centering
\includegraphics[width=0.942\columnwidth]{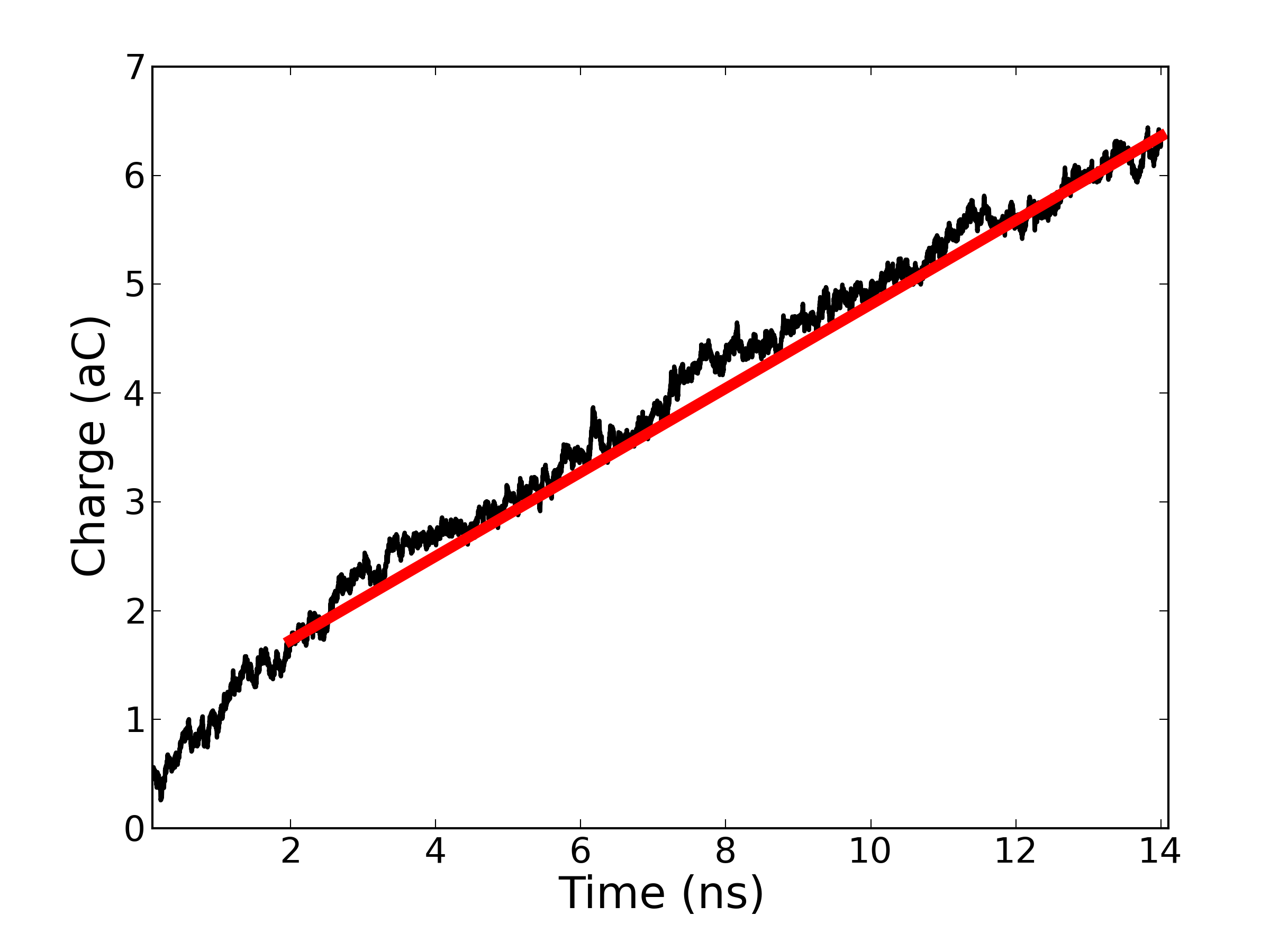}
\caption{\label{currentfig}(Color online) Typical plot of ionic current integrated over time, namely $\int_S dq$
where $S$ is the surface area of the transverse channel. The slope of this curve is the instantaneous current at each time in units of nA. We have used an interval of time $\Delta t=1$ ps. After about 2 ns a steady-state
current sets in. Its value is obtained from the slope of the red line obtained from the last 12 ns
of simulation. This particular plot is for poly(dA)$_7$.
}
\end{figure}
We calculate the current directly by looking at the motion of the ions in the system
 \begin{equation}I=\sum_{i}\frac{q_i\Delta z_i}{l_z\Delta t} \label{current:eq},  \end{equation}
where $l_z$ is the cell length of the system, and the sum is over all ions in the system. ${\Delta z_i}$ is the distance the $i$th ion has traveled in an interval of time $\Delta t$. Other methods of calculating the current were tested and they yielded similar results to equation (\ref{current:eq}). For instance, the slope of the curve $Q(t)=\int_S dq$ at steady state, where $S$ is the surface of the transverse channel (at steady state the choice of this surface is irrelevant~\cite{diventra:2008}) yields similar currents.

In about $1$ to $2$ ns, the ions move into a steady-state current-carrying state, so we do not begin to calculate the current until after this time period. The next 12 ns are then used to calculate the current for that run. This short transient is clear from Fig.~\ref{currentfig} where we plot $Q(t)$ for a typical run. Other, shorter, lengths of time were checked, but the best, most stable results were obtained with the present timescale. A rough estimate of the standard deviation of the current measurement is that it would scale as the inverse square root of the measurement window \cite{aksimentiev:2010}. Experimental measurements for a reasonable sampling frequency would average over a longer timescale than we have simulated, so we expect the current distributions to be even less noisy than we have calculated. Our rate of $12$ ns per sample would correspond to a sampling frequency of approximately $80$MHz. Ionic current measurements have been recently conducted at $1$MHz \cite{rosenstein:2012}, which would be a factor of $80$ slower than our simulations.
\begin{figure}[t]
\centering
\includegraphics[width=0.942\columnwidth]{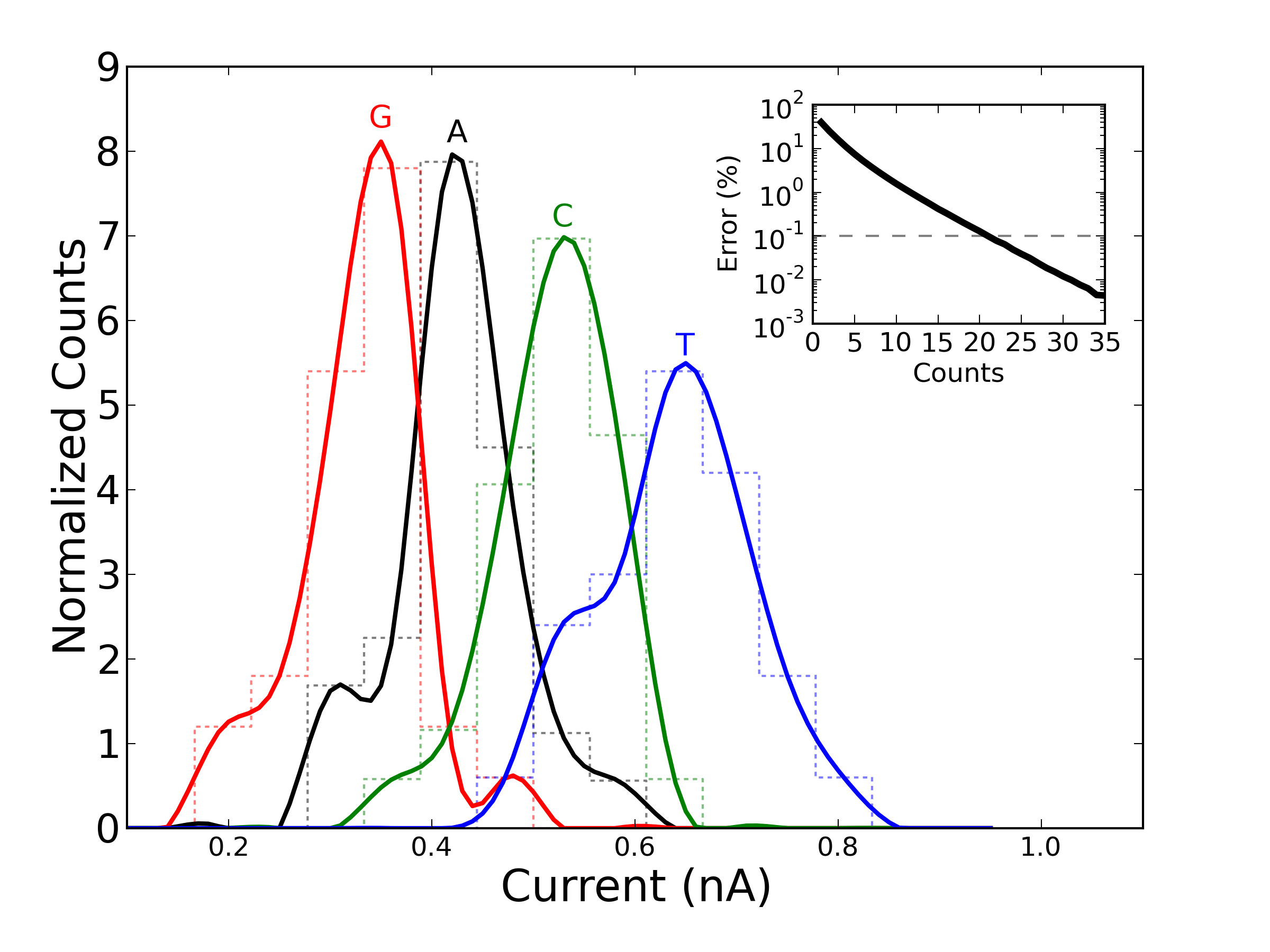}
\caption{\label{fig:distributions}(Color online) Transverse ionic current distributions for the different nucleotides in the pair of intersecting channels of Fig.~\ref{fig:Diagram}. The inset shows the probability of identifying a base incorrectly after measuring the current for that base "counts" times.
}
\end{figure}

The current distributions from the all-atom MD simulations are shown in Fig \ref{fig:distributions}. First we note, as was also done in the case of transverse electrical currents~\cite{lagerqvist:2006}, that one measurement of ionic current is not sufficient to sequence a DNA strand to high precision. However, the distributions we have obtained are sufficiently disjoint to not require a large number of measurements per base. We calculate below the number of measurements required based on our data, but it goes without saying that each device (in terms of channel widths, shape, etc.) would likely produce its own set of probability distributions.

Second, the average current we obtain suggests the order $\bar I_G<\bar I_A<\bar I_C<\bar I_T$. Apart from $C$ and $T$ this order is in reasonable agreement with the order of volumes of the different nucleotides $V_G>V_A>V_T>V_C$ \cite{zwolak:2008}, suggesting that the currents are somewhat correlated with the exclusion volume of each base \cite{zwolak:2008, branton:2008}. However, as also indicated by the reverse order of $C$ and $T$, the exclusion volume alone cannot fully explain the order of the average currents we have obtained, and other microscopic effects must also intervene such as the strength of the base dipoles, roughness of the surfaces at the channel intersection, and possibly other ones. Again, this confirms that the actual distributions will likely depend on the structural properties of each device.
\section{Analysis}
In order to sequence with this particular approach, we then parallel the protocol suggested in Ref.~\cite{lagerqvist:2006} by one of us (MD). First, run a strand of DNA with known composition through the longitudinal channel, e.g., a homogeneous strand of the four different bases. Then, while the DNA is translocating through the channel, measure the current as many times as possible and build up a distribution of currents for each DNA base. Once this distribution for each base is obtained, the DNA strand that is to be read is sent through. The current is read $N$ times for each base X as it passes through the intersection, and each of those $N$ readings is analyzed for the probability that it could be an $A$, $C$, $T$, or $G$. The probability of X being correctly identified after one measurement is given by~\cite{lagerqvist:2006}
\begin{equation} \langle P_X(I)\rangle =  \frac{n_X(I)}{n_A(I)+n_C(I)+n_T(I)+n_G(I)} \end{equation}
Where $n_X(I)$ is the height of the normalized distribution for current $I$, and $X$ is one of the bases $A, C, G, T$.

After $N$ measurements, the probability that the base at the intersection of the two channels is correctly identified is~\cite{lagerqvist:2006}
\begin{equation} \label{prob:EQ}
 \langle P_X^N\rangle=\frac{\prod\limits_i^N{n_X(I_i)}}{\sum\limits_{J=A,C,G,T}\left(\prod\limits_i^N{n_J(I_i)}\right)} \end{equation}
A Monte Carlo method was employed with the distribution data obtained from our simulations to calculate the probability of correctly identifying a base after $N$ measurements. For each base, $I_1$ through $I_N$ were randomly generated from the current distribution for that base. The probability $P_X$ is then calculated from equation (\ref{prob:EQ}). Assuming an equal proportion of each of the bases within the strand being sequenced, the probability of correctly identifying a base at random after $N$ independent measurements is just the average
\begin{equation} P_N=\frac14\left(P_A^N+P_C^N+P_G^N +P_T^N\right).\end{equation}
We then average over realizations to arrive at the probability, $P=\langle P_N\rangle$, that we correctly identify a random base after $N$ measurements.

In the inset of figure \ref{fig:distributions}, we plot a graph of uncertainty, $E=1-P$, in identifying bases versus number of independent measurements. When at least $20$ measurements are taken, the probability of error is less than $0.1\%$. This seems to suggest that, with these particular distributions, we need fewer measurements with this method than with transverse electronic transport for the same error rate~\cite{lagerqvist:2006}. However, as already noted ionic transport measurements are typically slower than electronic ones. Nevertheless, if we measure the ionic current at a reasonable rate of $100$kHz, then DNA could be identified at $5,000$ bases per second using this scheme. Therefore, an entire genome could be sequenced in less than 7 days without parallelization. Clearly, this estimate may be somewhat off according to the actual distributions obtained experimentally. Furthermore, the error rate assumes that the pore is made small enough that neighboring bases do not affect the current, and that it is easy to tell when a base passes over the intersection. If that is not the case, it may take several more measurements per base simply to know that a change in base at the intersection has occurred. These effects would all increase the number of measurements required.

\section{Conclusions}
We conclude that the approach we have described in this paper can indeed discriminate between the four DNA bases and has thus potential as an alternative sequencing method. Furthermore, this method would easily lend itself to parallelism. For example, a device dedicated to each chromosome would decrease the time required to sequence an entire genome by a factor of ten. Finally, this same idea could be implemented as a protocol for protein sequencing. We are now in the process of assessing the feasibility of this last possibility.

\noindent
{\bf Acknowledgment.} This research has been partially supported by the NIH-National Human Genome Research Institute. We also gratefully acknowledge discussions with M. Ramsey. Part of the calculations in this work were performed by using resources provided by the Open Science Grid, which is supported by the National Science Foundation and the U.S. Department of Energy.
\bibliographystyle{unsrt}
\bibliography{jawilson}

\end{document}